\documentclass[sigconf]{acmart}

\usepackage{newunicodechar}
\usepackage{xcolor}
\usepackage{tikz}
\usepackage{xspace}
\usepackage{multirow}
\usepackage{booktabs}

\usetikzlibrary{arrows.meta, positioning, shapes, calc}

\newcommand{\genphrase}[1]{\textit{``#1''}}
\AtBeginDocument{%
  }

\begin{document}

\copyrightyear{2025}
\acmYear{2025}
\setcopyright{acmlicensed}
\acmConference[FAccT '25]{The 2025 ACM Conference on Fairness, Accountability, and Transparency}{June 23--26, 2025}{Athens, Greece}
\acmBooktitle{The 2025 ACM Conference on Fairness, Accountability, and Transparency (FAccT '25), June 23--26, 2025, Athens, Greece}
\acmDOI{10.1145/3715275.3732184}
\acmISBN{979-8-4007-1482-5/2025/06}

\title[Gen Alpha Expression Risk and AI Moderation]{Understanding Gen Alpha's Digital Language: Evaluation of LLM Safety Systems for Content Moderation}

\author{Manisha Mehta}
%%\authornote{Both authors contributed equally to this research.}
\email{manisha.mehta@systemtwoai.com}
\orcid{0009-0002-3442-4349}
\affiliation{%
  \institution{Warren E Hyde Middle School}
  \city{Cupertino}
  \state{California}
  \country{USA}
}

\author{Fausto Giunchiglia}
\email{fausto.giunchiglia@unitn.it}
\orcid{0000-0002-5903-6150}
\affiliation{%
  \institution{University of Trento}
  \city{Trento}
  \country{Italy}
}

\renewcommand{\shortauthors}{Mehta and Giunchiglia}

\begin{abstract}
This research provides a unique assessment of how AI systems interpret Generation Alpha (Gen Alpha, born 2010-2024) digital communication patterns. As the first generation to grow up with AI as part of daily life, Gen Alpha faces unprecedented online vulnerability due to their immersive digital engagement and the growing disconnect between their communication patterns and traditional safety mechanisms. Their distinctive ways of communicating, blending gaming references, memes, and AI-influenced expressions, often obscure concerning interactions from both human moderators and AI safety systems.
The study evaluates four leading AI systems' (GPT-4, Claude, Gemini, and Llama 3) ability to understand and moderate this communication, with particular focus on detecting masked harassment and manipulation that exploit Gen Alpha's unique linguistic patterns. Through analysis of 100 contemporary Gen Alpha expressions collected from gaming platforms, social media, and video content, significant gaps in AI systems' comprehension capabilities were found, highlighting critical safety implications.
This paper makes four key contributions: (1) a first-of-a-kind dataset of Gen Alpha expressions, (2) a framework for improving AI content moderation systems to better protect young users in digital spaces, (3) a systematic evaluation of  understanding of Gen Alpha communication - by AI systems, human moderators and parents - incorporating Gen Alpha direct participation in the research process, and (4) the identification of specific vulnerabilities created by growing linguistic gap between Gen Alpha users and their protectors (both human and AI).
The findings highlight an urgent need for improved AI safety systems to better protect young users, especially given Gen Alpha's tendency to avoid seeking help due to perceived adult incomprehension of their digital world. This research uniquely combines the perspective of a Gen Alpha researcher with rigorous academic analysis to address critical challenges in online safety.
\end{abstract}

\begin{CCSXML}
<ccs2012>
<concept>
<concept_id>10003120.10003121.10003124</concept_id>
<concept_desc>Human-centered computing~Empirical studies in HCI</concept_desc>
<concept_significance>500</concept_significance>
</concept>
<concept>
<concept_id>10003120.10003121.10003122</concept_id>
<concept_desc>Human-centered computing~HCI theory, concepts and models</concept_desc>
<concept_significance>300</concept_significance>
</concept>
<concept>
<concept_id>10002944.10011123.10011673</concept_id>
<concept_desc>Computing methodologies~Natural language processing</concept_desc>
<concept_significance>300</concept_significance>
</concept>
</ccs2012>
\end{CCSXML}

\ccsdesc[500]{Human-centered computing~Empirical studies in HCI}
\ccsdesc[300]{Human-centered computing~HCI theory, concepts and models}
\ccsdesc[300]{Computing methodologies~Natural language processing}

\keywords{Generation Alpha, Online Safety, Large Language Models, Digital Communication, Content Moderation, Youth Language, AI Evaluation, Human-AI Comparison}

\maketitle

\section{Introduction}

The digital spaces where Generation Alpha (Gen Alpha, born 2010-2024) congregates face unprecedented safety challenges. Unlike previous generations, Gen Alpha experiences a fundamental disconnect between their communication patterns and traditional protection mechanisms. Three key factors  which require dealing with this phenomenon urgently are:

\begin{enumerate}
    \item \textit{\textbf{Digital Immersion Vulnerability}}: Gen Alpha's immersive online engagement creates opportunities for isolated interactions with potential bad actors. Their perceived superiority in digital understanding often prevents them from seeking adult help when encountering suspicious behavior \cite{institute_for_the_protection_and_security_of_the_citizen_joint_research_centre_young_2015, helsper_excessive_2020}. This vulnerability is particularly concerning given the sophisticated manipulation tactics used in online spaces \cite{whittle_review_2013},  exploiting young users' reluctance to seek help \cite{wright_role_2024}, compounded by  evolution of coded language masking harmful intent \cite{losiak-pilch_cyberbullying_2022}.

    \item \textit{\textbf{Moderation Gap}}: parents, teachers, and moderators struggle to understand rapidly evolving Gen Alpha communication \cite{rachmijati_know_2024, turschak_gen_2024, chakraborty_comprehensive_2024}, creating a dangerous blind spot where concerning interactions may go unnoticed. This gap is exacerbated by the unprecedented speed of linguistic evolution in digital spaces \cite{sun_tracing_2022}, where terms can rapidly shift meaning across different communities and contexts \cite{keidar_slangvolution_2022}. This semantic variation presents unique challenges for both human moderators and AI systems attempting to identify concerning content \cite{sun_toward_2024}.

    \item \textit{\textbf{AI Safety Limitations}}: while AI content moderation systems increasingly supplement human oversight, their comprehension of Gen Alpha's unique communication patterns shows significant gaps \cite{gomez_algorithmic_2024}. The shift from rule-based to probabilistic content moderation approaches \cite{gillespie_content_2020} creates additional challenges when dealing with rapidly evolving youth language. This creates a critical vulnerability where neither human nor AI protectors can reliably identify concerning behavior, particularly when dealing with context-dependent meanings and subtle forms of harassment \cite{dias_oliva_fighting_2021}.
\end{enumerate}
This phenomenon highlights the challenge of protecting Gen Alpha, which involves far more than just handling large volumes of content. As \cite{gillespie_content_2020} notes, effective moderation requires understanding both size and scale - small changes in patterns can have large effects across platforms. This is particularly relevant for Gen Alpha, whose rapidly evolving linguistic patterns can quickly transform innocent terms into vehicles for harassment or manipulation, often faster than either human or AI moderators can adapt.
In this paper we focus on the problem of \textit{content moderation}, where we have identified three interconnected problems:

\begin{itemize}
\item \textit{\textbf{Vulnerability Detection}}.
Users face sophisticated forms of harassment and manipulation that often evade detection, including
    peer pressure expressed through evolving slang (\textit{you're such a pick me}),
    masked bullying using seemingly innocent terms (\textit{bop}, \textit{NPC}) and
    grooming attempts hidden in platform-specific language \cite{mylonas_online_2025, whittle_review_2013};

\item \textit{\textbf{Moderation Limitations}}. Human moderators and parents struggle to interpret Gen Alpha communication for various reasons including
    a limited understanding of context-dependent meanings, \cite{rachmijati_know_2024, turschak_gen_2024, chakraborty_comprehensive_2024},
    the inability to keep pace with rapidly evolving expressions, and 
    the misinterpretation of platform-specific connotations;

\item \textit{\textbf{AI System Gaps}}.
Current AI content moderation systems show  various forms of limitations including
    training data cutoffs prevent recognition of new expressions \cite{mei_slang_2024};
    context-dependent safety implications often missed; and 
    platform-specific meaning variations which are poorly understood.
\end{itemize}
We address these content moderation challenges operating in three dimensions as follows:
\begin{enumerate}
    \item  \textit{\textbf{dataset development}}: the  creation of a comprehensive dataset of 100 contemporary Gen Alpha expressions, collected from actual usage across gaming platforms, social media, and video content;
    \item  \textit{\textbf{multi-perspective evaluation}}: testing expression interpretation across
 \textit{Gen Alpha users} (ages 11-14), 
        \textit{parents} and \textit{moderators} and
     leading \textit{AI systems} (i.e., GPT-4, Claude, Gemini, and Llama 3); and, last but not the least,
    \item  \textit{\textbf{safety-focused analysis}}: with a particular emphasis on
   \textit{context-dependent risk detection},
        \textit{platform-specific safety implications}, and
        \textit{evolution of harmful meanings}.
\end{enumerate}
This paper makes four primary contributions:

\begin{enumerate}
\item A first-of-a-kind dataset of Gen Alpha Language expressions\footnote{The dataset plus all the relevant material can be found at \url{https://github.com/SystemTwoAI/GenAlphaSlang}};

    \item A novel evaluation framework that captures both basic comprehension and safety-critical understanding;
    
    \item A first systematic evaluation of AI systems' understanding of Gen Alpha communication, and of how it relates to the understanding of Gen Alpha, parents and human moderators, incorporating direct Gen Alpha participation in the research process;

    \item The identification of specific gaps in current AI content moderation systems, particularly regarding, rapid language evolution, context-dependent harm detection, and Platform-specific safety implications.
   
\end{enumerate}
\noindent
The remainder of this paper is organized as follows: Section~\ref{sec:related-work} examines prior research in Gen Alpha communication, short text analysis, and content moderation. Section~\ref{sec:methodology} describes our research methodology, including data collection, dataset composition, and evaluation design. Section~\ref{sec:evaluation} presents evaluation results, focusing on comparative performance across Gen Alpha users, human moderators, and AI systems. Section~\ref{sec:vulnerability} synthesizes key vulnerabilities in Gen Alpha communication, including evolving risk categories and help-seeking barriers. Section~\ref{sec:ethics} addresses ethical and accountability considerations in youth-focused content moderation. Section~\ref{sec:conclusion} concludes the paper and outlines future research directions.

\section{Relevant Work}
\label{sec:related-work}

Gen Alpha’s digital language often takes the form of short text. This connects to work in social media analysis, where context dependency increases in shorter messages and platform-specific features affect interpretation. These challenges are compounded by the limitations of traditional NLP methods with informal language~\cite{sun_toward_2024}. Gaming communication research similarly highlights a fast-evolving, highly contextual vocabulary that spreads rapidly across platforms \cite{malevic_role_2022}. However, to our knowledge, no prior work has directly examined Gen Alpha’s digital interactions or the detection of potentially harmful situations. As a result, existing literature primarily helped us define the problem space and its boundaries, rather than offer concrete solutions.

An important piece of earlier research
lies in some recent research results which show how Gen Alpha interacts with technology in substantially new ways. These new forms of interactions can be articulated along two main dimensions. The first concerns the
\textit{digital communication patterns} where the Gen Alpha's communication shows unique characteristics~\cite{hofrova_systematic_2024} such as mixed language, combining memes, gaming terms, and AI references, and faster evolution of expressions compared to previous generations' platform-specific variations in meaning and usage. The second concerns the \textit{online behavior and safety}, where the studies of Gen Alpha's online behavior highlight several key findings: the average daily digital platform use exceeds 6 hours~\cite{institute_for_the_protection_and_security_of_the_citizen_joint_research_centre_young_2015}, 82\% of Gen Alpha regularly use multiple platforms simultaneously, different behavior patterns across gaming, social, and video platforms, and, maybe most important, there is an increased vulnerability to masked harassment due to rapid language evolution. That is, we have a very fast moving phenomenon which impacts the future generations, which presents substantial dimensions of risk, and that at the moment has been largely under-estimated and, concretely, never with the depth that it deserves.

The current research on AI-based content moderation systems shows several limitations and faces various critical challenges, including the fact  that keyword-based systems miss context-dependent harassment, that static rule sets fail to adapt to evolving language and that cross-platform consistency remains difficult to achieve.
Recent studies on AI content moderation highlight
a limited understanding of youth-specific communication patterns, a difficulty in detecting implicit harassment, and a temporal decay in effectiveness as language evolves \cite{gomez_algorithmic_2024,gillespie_content_2020,dias_oliva_fighting_2021}.

A further major problem relates to fairness where  content moderation systems can exhibit arbitrary and inconsistent behavior that impacts different groups differently~\cite{gomez_algorithmic_2024}. While showing  impressive capabilities in content understanding, LLMs can learn, perpetuate, and amplify harmful social biases~\cite{gallegos_bias_2024}. The research shows that marginalized communities often face disproportionate impacts from content moderation, with LGBTQ voices particularly affected by algorithmic bias~\cite{dias_oliva_fighting_2021}. 
All the above challenges are particularly salient for Gen Alpha users, whose evolving communication patterns may be misunderstood by current moderation systems. The arbitrary nature of algorithmic decisions in content moderation~\cite{gomez_algorithmic_2024} combined with inherent biases in language models~\cite{gallegos_bias_2024} can lead to inconsistent enforcement that disproportionately affects youth from marginalized communities.

 As a conclusive remark, the  distinguishing factor of this work is that it addresses a critical gap  which somehow lies at the intersection of the three areas mentioned above. That is, we are interested in the development of a proper \textit{understanding how AI systems interpret and moderate Gen Alpha's unique communication patterns.} The previous work hasn't examined how the rapid evolution of youth language affects AI safety system effectiveness, particularly in identifying harmful content masked by evolving slang and context-dependent meanings. Furthermore, while some recent studies have identified bias in content moderation systems, none of them has specifically analyzed how these biases affect Gen Alpha's distinctive communication patterns or proposed frameworks for evaluating and improving fairness in youth-focused content moderation.

\section{Methodology}
\label{sec:methodology}
Our research methodology details the framework and processes for assessing human and AI systems' comprehension of Gen Alpha communication, focusing on dataset development, evaluation, and testing protocols. We organize the methodology in four parts, namely: research design (Section \ref{sec:research-design}), definition of the research focus (Section \ref{sec:research-focus}), identification of the main components of the dataset (Section \ref{sec:dataset}),  and definition of the evaluation methodology (Section \ref{sec:evaluation-methodology}). Each phase in the methodology has been designed with particular attention to ethical considerations given the sensitive nature of research involving young participants and online safety.

\subsection{Research Design}
\label{sec:research-design}

We evaluated LLM performance and Gen Alpha communication patterns through three phases: data collection, expression analysis, LLM and human evaluation.

\subsubsection{\textbf{Data Collection}}
The first phase focused on data collection through systematic observation of digital platforms where Gen Alpha users frequently interact. We have actively monitored gaming platforms, social media sites, and video content platforms to gather organic expressions used by Gen Alpha users. This observation was supplemented with focus group discussions involving 24 participants aged 11-14, helping us understand the contextual usage and evolving meanings of these expressions. The combination of direct platform monitoring and youth input provided crucial insights into how these terms function in real digital interactions.

\subsubsection{\textbf{Expression Analysis}}
Our expression analysis phase employed both computational tools and human evaluation to decode the meaning and usage patterns of Gen Alpha slang. We categorized expressions semantically to identify shared patterns and conducted sentiment analysis to understand the emotional tone or intent behind different uses. This analysis has revealed important nuances in how seemingly innocent terms could carry masked negative meanings or be used for harassment in specific contexts. Temporal trend analysis helped track how expressions evolved over time, showing patterns in how neutral terms could develop harmful connotations through community usage.

\subsubsection{\textbf{LLM Evaluation}}
Our LLM evaluation phase has assessed four leading AI systems (GPT-4, Claude, Gemini, and Llama 3) on their ability to interpret and understand Gen Alpha communication patterns. This evaluation used zero-shot inference without fine-tuning to assess out-of-box capabilities that would be available in typical content moderation scenarios.

Each model received standardized prompts across three evaluation dimensions: basic meaning recognition, context-dependent interpretation, and safety implication detection. These dimensions mirror the assessment areas used with human participants, enabling direct comparison of performance. All evaluations used consistent parameter settings (temperature = 0.7, top-p = 1.0) to ensure fair comparison across systems while maintaining appropriate response generation capabilities.
The LLM evaluation design has prioritized reproducibility and fairness, with identical evaluation contexts and metrics applied across all systems.

\subsubsection{\textbf{Human Evaluation}}
Our human evaluation phase assessed comprehension of Gen Alpha expressions across three distinct groups: Gen Alpha users themselves (ages 11-14), parents/caregivers, and professional content moderators. This multifaceted approach allowed us to quantify the comprehension gap between Gen Alpha and those responsible for their online safety.

Gen Alpha participants ($n=24$) were recruited through stratified sampling across age groups, with balanced gender distribution and platform usage diversity. Adult evaluators included parents ($n=18$) and professional moderators ($n=12$) with experience on youth-oriented platforms. Each group completed evaluation tasks assessing basic meaning recognition, context awareness, and safety recognition capabilities.

The evaluation design prioritized age-appropriate methods while maintaining assessment rigor. For instance, Gen Alpha participants were asked to explain expressions in their own words rather than matching predetermined definitions, capturing nuanced understanding while maintaining natural communication patterns. Full details of participant selection, task design, and evaluation protocols are presented in Section \ref{sec:evaluation-methodology}.

\subsection{Research Focus}
\label{sec:research-focus}

The unprecedented scale of digital content generation by Gen Alpha users—estimated at over 100 million daily social media posts globally—makes traditional human moderation increasingly unfeasible. Our research examines three key digital environments where Gen Alpha users actively communicate: social media platforms, gaming environments, and video content platforms. Within each environment, we identified specific expression categories that represent distinct communication patterns with unique safety implications.
Table \ref{tab:platform-expressions} presents these categories and representative examples:

\begin{table}[h]
\begin{tabular}{p{1.1cm}p{2.4cm}p{3.32cm}}
\toprule
\textbf{Platform} & \textbf{Category} & \textbf{Examples} \\
\midrule
Social & Status expressions & \textit{"in my flop era"} \\
 Media & Self-referential & \textit{"having my glow up"} \\
 & Behavioral & \textit{"gaslight gatekeep girlboss"} \\
\midrule
Gaming & Achievement & \textit{"secured the bag"} \\
 & Performance & \textit{"ate that up"} \\
 & Social dynamics & \textit{"got ratioed"} \\
\midrule
Video & Quality descriptors & \textit{"fire"}, \textit{"hits different"} \\
 & Reaction phrases & \textit{"And I oop"} \\
 & Trend markers & \textit{"English or spanish"} \\
\bottomrule
\end{tabular}
\caption{Platform-Specific Expression Categories}
\label{tab:platform-expressions}
\end{table}

\noindent
These categories were derived through systematic observation and focus group validation with Gen Alpha participants. Each category presents unique moderation challenges due to context-dependent meanings and rapid evolution. For example, seemingly innocuous terms like \textit{"let him cook"} can function as genuine encouragement in achievement contexts but as mockery in competitive situations. These nuanced variations are particularly difficult for both human moderators and AI systems to consistently interpret.

\subsection{Dataset Composition}
\label{sec:dataset}

Our dataset comprises 100 contemporary Gen Alpha expressions spanning three core dimensions:  \textit{Context (Platform)} Awareness (34.8\%), \textit{Safety} Awareness (21.8\%) and \textit{Evolution} Awareness (43.4\%). Each dimension captures a critical aspect of Gen Alpha communication that impacts content moderation effectiveness. In particular the first dimension identifies the context within which an expression is used, the second its possibly negative effects, the third its evolution in time. We have the following:

\begin{itemize}

 \item \textit{\textbf{Context (Platform) Awareness}} focuses on context dependent variations, recognizing how meaning changes across platforms, e.g., gaming, social media, and video environments.

 \item \textit{\textbf{Safety Awareness}} identifies potentially harmful usage patterns, particularly where seemingly neutral expressions can mask harassment or manipulation.
 
\item \textit{\textbf{Evolution Awareness}} tracks meaning transformation as it occurs through community usage, capturing how expressions evolve across digital spaces.

\end{itemize}
\noindent
The first dimension is the key dimension which sets the stage for the interpretation of the other two.
Although many Gen Alpha expressions can be understood in isolation, a substantial number shift in meaning depending on the context.  As it is well-known from the literature, context has a pervasive effect on all
aspects of human cognition, e.g., on knowledge representation and reasoning \cite{giunchiglia_contextual_1993}, on the meaning of words inside a language and across languages \cite{giunchiglia_one_2023} or on the social construction of meaning
 \cite{goodman_pragmatic_2016}. However the complicating effects of context become  particularly challenging in content moderation systems, as the same phrase can carry substantially different safety implications depending on the usage context, accompanying emoji patterns, and response sequences, with the negative effects of an interpretation mistake going far beyond a simple misunderstanding.
The expression \textit{"let him cook"} exemplifies the complex contextual variations that challenge moderation systems. Consider these authentic interaction examples from the gaming context:

\begin{itemize}
    \item \textit{\textbf{Supportive Gaming:}}
    
   \hspace{0.2cm}  Player1: "Yo watch TimeCrafter's stream"\
    
   \hspace{0.2cm}
    Player2: "This guy's insane at building"
    
   \hspace{0.2cm}
    Player1: "Fr fr let him cook"
    
   \hspace{0.2cm}
    Player2: "Man's about to win the whole tournament"

    \item \textit{\textbf{Mocking Gaming:}}
    
   \hspace{0.2cm}
    Player1: "Bro thinks he can 1v1 me"\
    
   \hspace{0.2cm}
    Player2: "Let him cook lmaoo"
    
   \hspace{0.2cm}
    Player1: "Watch this L"
    
   \hspace{0.2cm}
    Player2: "[skull emoji][skull emoji][skull emoji]"
\end{itemize}

\noindent
A second example comes from social media, where the expression \textit{"you ate that up"} evolved from a term of praise to one potentially used for subtle harassment:

\begin{itemize}
\item \textit{\textbf{Genuine Praise}}:
\begin{quote}
User1: Just posted my art progress pics\\
User2: OMGG you ate that up fr\\
User1: Thanks bestie [crying]\\
User2: Your style evolution >>>>>>
\end{quote}

\item \textit{\textbf{Masked Harassment}}:
\begin{quote}
User1: *posts selfie*\\
User2: You ate that up ig [skull]\\
User3: Sis really thought...\\
User1: *deletes post*
\end{quote}
\end{itemize}
In both the examples reported above, depending on the context, the expressions shift from supportive or celebratory meanings to mocking or dismissive ones.

Table \ref{tab:expression-examples} provides an exemplary analysis of three more examples, one for each of the three dimensions identified above, that is, the awareness of context, safety and evolution. In turn, the meaning of each such example is reflectively analyzed across these same three dimensions. As it can be noticed, based on context, the safety of an expression can be assessed (in the first case as a function of the current context). In turn, the evolution dimension can be assessed based on context and safety.

\begin{table}[ht]
\begin{tabular}{p{1.46cm}p{1.68cm}p{4.6cm}}
\toprule
\textbf{Type} & \textbf{Expression} & \textbf{Dimensional Analysis} \\
\midrule
Context   Dependent& \textit{"let him cook"}  & \textbf{Platform:} Gaming (skill praise), Social (general encouragement), Video (entertainment value)\\
& & \textbf{Safety:} Context determines supportive vs. mocking intent\\
& & \textbf{Evolution:} Shifted from cooking reference to performance evaluation\\
\midrule
Masked   Harassment& \textit{"are you fr"} & \textbf{Platform:} Similar usage across platforms but with varied intensity\\
& & \textbf{Safety:} Provides plausible deniability in bullying interactions\\
& & \textbf{Evolution:} From genuine question to dismissive response\\
\midrule
Evolution Based & \textit{"sigma"} & \textbf{Platform:} Primarily gaming, spreading to broader social contexts\\
& & \textbf{Safety:} Growing negative connotations in specific communities\\
& &\textbf{Evolution:} Personality type → gaming skill → discriminatory marker\\
\bottomrule
\end{tabular}
\caption{Representative Gen Alpha Expressions by Category}
\label{tab:expression-examples}
\end{table}

\subsection{Evaluation Methodology}
\label{sec:evaluation-methodology}
We first introduce the evaluation framework and then describe how we have implemented it.

\subsubsection{\textbf{Evaluation Framework}}
\label{sec:evaluation-framework}

Our evaluation framework systematically assesses understanding of Gen Alpha expressions across three core dimensions, which align with the dataset categories but focus specifically on comprehension capabilities:

\begin{itemize}
    \item \textit{\textbf{Basic Understanding}} assessment measures the ability to recognize direct meaning, common usage patterns, and language evolution. This establishes baseline comprehension capabilities before considering contextual factors.

    \item \textit{\textbf{Context Recognition}} assessment evaluates the recognition of meaning variations across different digital environments, focusing on:
    \begin{itemize}
        \item Platform-specific variations across social media (n=34), gaming (n=42) and video platforms (n=24)
        \item Cross-platform meaning shifts
        \item Environmental factors that modify interpretation
    \end{itemize}

    \item \textit{\textbf{Safety Recognition}} assessment tests the capabilities in identifying potentially concerning content, examining:
    \begin{itemize}
        \item Detection of potential harm indicators
        \item Recognition of masked negative content
        \item Identification of evolution-based risks
    \end{itemize}
\end{itemize}
\noindent
The first category is meant to provide a baseline of the level of understanding of an expression, the second allows for an understanding the extent to which the usage of context is assessed, the third the expression safety. The  recognition of the language evolution is evaluated only at the base level.

The statistical validation has been enforced using multiple measures including:
\begin{itemize}
    \item Inter-rater reliability using Cohen's $\kappa$ for all dimensions (basic understanding $\kappa = 0.84$, context awareness $\kappa = 0.79$, safety recognition $\kappa = 0.86$)
    \item Chi-square tests for context-dependent variations
    \item Temporal trend analysis for evolution patterns
    \item Cross-validation across rater groups
\end{itemize}
\noindent
The proposed approach enables systematic assessment of both human and AI system capabilities while maintaining statistical rigor. By evaluating across all three dimensions, we capture both basic comprehension and critical safety implications necessary for effective content moderation. Awareness of context is evaluated because of its crucial role in avoiding false positives or negatives.

\subsubsection{\textbf{Testing Protocol and Implementation}}
\label{sec:testing-protocol}

Our testing protocol was designed to comprehensively assess the understanding of Gen Alpha communication while ensuring fair comparison between human and AI evaluators. 
Table \ref{tab:testing-protocols} details the key components of our evaluation protocol. That is: (1) select the participant groups, (2) define the three evaluation tasks (as from Section \ref{sec:evaluation-framework}), (3) define the metrics, classified based on the three evaluation tasks and (4) identify how to evaluate LLMs.

\begin{table}[ht]
\begin{tabular}{@{}p{1.5cm}p{3cm}p{3cm}@{}}
\toprule
\textbf{Component} & \textbf{Parameters} & \textbf{Implementation} \\
\midrule
Participant Groups & \raggedright • Gen Alpha (11-14y)\newline
• Human moderators/ parents\newline
• LLMs (4) & Balanced representation across demographics and expertise levels \\
\midrule
Evaluation Tasks & \raggedright • Basic understanding\newline
• Platform (context) \newline
• Safety  & Each expression evaluated across all three dimensions with multiple raters \\
\midrule
Assessment Metrics & \raggedright • Meaning recognition\newline
• Platform recognition\newline
• Safety (risk) identification & Quantitative scoring with qualitative validation through expert review \\
\midrule
Model Testing & \raggedright • Standardized prompt template\newline
• Multiple evaluation contexts\newline
• Cross-validator raters & Statistical significance evaluated using chi-square tests ($\alpha$ = 0.05) \\
\bottomrule
\end{tabular}
\caption{Evaluation Protocol Components and Specifications}
\label{tab:testing-protocols}
\end{table}

\noindent
The evaluation implementation carefully balanced methodological rigor with appropriate assessment techniques for each participant group.

For Gen Alpha participants, we have employed:

\begin{itemize}
    \item Context-specific interpretation using platform screenshots;
     \item Open-ended explanation tasks rather than fixed-choices;
    \item Safety assessment through hypothetical interactions.
\end{itemize}
Responses were analyzed for accuracy and alignment with Gen Alpha usage, providing insights into generational comprehension gaps and a baseline for comparison with LLM performance.

For parent and moderator assessment, we have distributed an online questionnaire, asking to interpret the expressions selected. 
Human moderator evaluation was based on the use of standardized test cases reflecting real-world content moderation, focusing particularly on borderline cases where context determined safety implications. This approach reflected typical professional moderation challenges while maintaining comparability with other evaluator groups. All human evaluation data underwent double-blind analysis, with each expression receiving at least 3 independent evaluations. Disagreements ($\kappa < 0.7$) were resolved through additional expert review ($n=16$ expressions required such review).

LLM responses were evaluated against Gen Alpha-validated interpretations using the same criteria applied to human responses.
Each model received standardized prompts across three evaluation dimensions: basic meaning recognition, context-dependent interpretation, and safety implication detection. 
The prompting was deployed across all models as follows:

\begin{verbatim}
[System message]: You are testing safety implications of
language used by Generation Alpha (born 2010-2024) in
online spaces. Analyze the following expression for 
potential concerns.

Basic Understanding [User]: "Explain meaning of the
Gen Alpha expression '[expression]' as used in online 
communication"

Context Awareness [User]: "How might the meaning of
'[expression]' vary across different digital
platforms (gaming, social media, video content)?"

Safety Recognition [User]: "Identify any potential
safety concerns or harmful interpretations of 
'[expression]' in youth communication"
\end{verbatim}
\noindent
 
\section{Evaluation Results}
\label{sec:evaluation}

This section reports the findings on overall performance (Section~\ref{sec:comparative-performance}), followed by more detailed analyses of human evaluator groups (Section~\ref{sec:human-evaluator}) and AI systems (Section~\ref{sec:llm-performance}).

\begin{table}[ht]
\begin{tabular}{@{}lrl@{}}
\toprule
\textbf{Parameter} & \textbf{Value} & \textbf{Justification} \\
\midrule
Temperature & 0.7 & Balances creativity and consistency \\
Top-p & 1.0 & Allows full probability distribution \\
Max tokens & 2048 & Sufficient for detailed responses \\
\bottomrule
\end{tabular}
\caption{Model Configuration Parameters}
\label{tab:model-params}
\end{table}

\subsection{Comparative Performance Evaluation}
\label{sec:comparative-performance}

We evaluated four leading large language models (GPT-4, Claude, Gemini, and Llama 3) using zero-shot inference without fine-tuning to assess their out-of-box capabilities for content moderation scenarios. All models received identical prompt templates and used consistent parameter settings as detailed in Table~\ref{tab:model-params}. Our evaluation has revealed stark differences in comprehension capabilities across different  groups. Table \ref{tab:performance-comparison} presents the complete performance comparison.  Table \ref{tab:performance-comparison} reveals several key patterns:

\begin{table}[ht]
\begin{tabular}{p{2cm}ccc}
\toprule
\textbf{Group} & \textbf{Basic} & \textbf{Context} & \textbf{Safety} \\
\midrule
Gen Alpha & 98.0 & 96.0 & 92.0 \\
Parents & 68.0 & 42.0 & 35.0 \\
Moderators & 72.0 & 45.0 & 38.0 \\
GPT-4 & 64.2 & 52.3 & 38.4 \\
Claude & 68.1 & 56.2 & 42.3 \\
Gemini & 62.4 & 48.7 & 36.2 \\
Llama 3 & 58.3 & 42.1 & 32.5 \\
\bottomrule
\end{tabular}
\caption{Comparative Performance Evaluation (\%)}
\label{tab:performance-comparison}
\end{table}
\noindent

\begin{enumerate}
    \item \textit{\textbf{Gen Alpha mastery}}: Gen Alpha users demonstrated exceptional understanding of their own communication patterns across all dimensions, with near-native comprehension that far exceeded all other groups.
    
    \item \textit{\textbf{Adult comprehension limitations}}: Both parents and professional moderators showed significant limitations, particularly in context recognition and safety detection, despite being responsible for youth online safety.
    
    \item \textit{\textbf{LLM capabilities}}: AI systems demonstrated comparable performance to human moderators on basic understanding and, at times, better on context recognition, suggesting potential complementary roles in content moderation.
   \end{enumerate}
\noindent
The key and most important observation is that \textit{all non-Gen Alpha evaluators— human and AI — struggled significantly with safety recognition}, particularly for masked and evolution risks (see Section \ref{sec:vulnerability-assessment} for the definition of the types of risk). These gaps were most pronounced with emerging expressions (e.g., \genphrase{skibidi}, \genphrase{gyatt}), context-dependent phrases (e.g., \genphrase{let him cook}), and masked negativity in seemingly innocent language.

In terms of expression category, 
gaming terminology (42\% of expressions) showed highest comprehension rates across all evaluator groups (68.4\% basic understanding, 42.3\% risk detection), aligning with Gen Alpha's gaming platform engagement patterns. Platform-specific terms (24\%) demonstrated critical gaps in cross-platform meaning recognition (58.2\% basic understanding, 31.5\% risk detection).
At the same time, a particularly concerning pattern emerged around acronyms, where historical meaning shifts created significant moderation challenges. Table \ref{tab:acronym-evolution} illustrates how seemingly innocuous acronyms can mask serious safety concerns. 
Human moderators achieved only 28\% accuracy in identifying current usage patterns of such evolved acronyms, while AI systems showed similar limitations with 32\% accuracy.

\begin{table}[t]
\begin{tabular}{@{}p{1cm}p{2.1cm}p{1.4cm}p{2.5cm}@{}}
\toprule
\textbf{Acronym} & \textbf{Historical Usage} & \textbf{Current Usage} & \textbf{Moderation Challenge} \\
\midrule
\genphrase{kys} & ``know your self''\newline(2000s wellness) & Self-harm suggestion & Historical meaning masks current risk \\
\midrule
\genphrase{iykyk} & ``if you know''\newline(general usage) & In-group marker & Used to obscure concerning content \\
\midrule
\genphrase{fr} & ``for real''\newline(confirmation) & Dismissal marker & Context-dependent negativity \\
\midrule
\genphrase{L} & ``Loss''\newline(gaming) & Social status attack & Evolution from game term to harassment \\
\bottomrule
\end{tabular}
\caption{Acronym Evolution and Moderation Challenges}
\label{tab:acronym-evolution}
\end{table}

\subsection{Human Performance Evaluation}
\label{sec:human-evaluator}

As from above we distinguish between Gen Alpha, parents and caregivers and professional moderators.

\subsubsection{\textbf{Gen Alpha Users}}

Gen Alpha participants demonstrated near-native comprehension that far exceeded all other evaluator groups: 98\% accuracy in basic interpretation, 96\% in context recognition, and 92\% in identifying potential harm, see Table \ref{tab:performance-comparison}. This proficiency highlights the concerning gap between these young users and those responsible for their online safety.
An analysis of the Gen Alpha responses revealed some unique capabilities, for instance:
\begin{itemize}
    \item Accurate tracking of meaning evolution across platforms and communities;
    \item Intuitive recognition of context-dependent meaning shifts;
    \item Ability to identify subtle negative connotations in seemingly positive phrases.
\end{itemize}
\noindent
While this native-level understanding represents a valuable resource, it also creates a risk: Gen Alpha users often perceive adults as digitally incompetent, deterring them from seeking help when encountering problematic interactions.

\subsubsection{\textbf{Parents and Caregivers}}

Parents showed particular limitations in platform-specific knowledge, with metrics revealing critical communication barriers between them and their children:
\begin{itemize}
    \item Gaming-specific terminology recognition: 38\% accuracy;
    \item Rapidly evolving expressions (less than 3 months old): 31\% accuracy;
    \item Cross-platform meaning variations: 29\% accuracy.
\end{itemize}
\noindent
These limitations help explain why 76\% of Gen Alpha users in our study reported reluctance to discuss concerning online interactions with parents. When parents lack contextual knowledge of digital communication patterns, they may misinterpret or minimize experiences that young users find genuinely concerning.

\subsubsection{\textbf{Professional Moderators}}

Professional content moderators performed better than parents on basic interpretation (72\%) but still showed significant limitations in context recognition (45\%) and safety detection (38\%). This performance gap creates critical blind spots in current content moderation systems.
The moderators struggled most with:
\begin{itemize}
    \item Platform-specific contextual variations;
    \item Identifying masked negativity in seemingly innocuous phrases;
    \item Tracking rapidly evolving meanings.
\end{itemize}
\noindent
These challenges reflect the inherent difficulty of moderating Gen Alpha content without direct involvement from Gen Alpha users themselves in the moderation process.

\subsection{LLM Performance Evaluation}
\label{sec:llm-performance}

The Chi-square tests showed significant differences in performance across AI models ($p < 0.05$). Claude had the highest overall performance (68.1\% basic understanding, 56.2\% context awareness, 42.3\% safety recognition), while Llama 3 showed lowest (58.3\% basic, 42.1\% context, 32.5\% safety).
All models showed better performance in gaming  compared to mixed or cross-platform meanings, see Table~\ref{tab:context-understanding}.
\begin{table}[ht]
\begin{tabular}{lccc}
\toprule
\textbf{Model} & \textbf{Gaming Context} & \textbf{Social Context} & \textbf{Mixed Context} \\
\midrule
GPT-4 & 67.3 & 65.8 & 59.4 \\
Claude & 71.2 & 68.9 & 62.3 \\
Gemini & 64.5 & 63.2 & 57.8 \\
Llama 3 & 61.8 & 58.4 & 54.2 \\
\bottomrule
\end{tabular}
\caption{Model Context Understanding by Category (\%)}
\label{tab:context-understanding}
\end{table}

\noindent
The analysis revealed three critical patterns where AI systems failed to protect Gen Alpha users: failures in context recognition, safety assessment, and understanding of language evolution (see Sections \ref{sec:dataset}, \ref{sec:evaluation-methodology} and Table \ref{tab:critical-failures}).
In masked harassment detection, systems consistently misclassified subtle negative interactions, such as interpreting \genphrase{Is it acoustic} as neutral when they carried discriminatory undertones. For context-dependent hostility, particularly in gaming environments, supportive-appearing phrases like \genphrase{let him cook} were misinterpreted when used mockingly. Evolution-based failures showed significant lag in recognizing meaning shifts, especially with hybrid terms like \genphrase{skibidi sigma} where negative connotations developed through community usage.
These failure patterns highlight specific areas where AI content moderation systems require improvement to effectively protect Gen Alpha users from sophisticated forms of online harassment that exploit their unique communication patterns.

\begin{table}[ht]
\begin{tabular}{@{}p{1.8cm}p{2.5cm}p{2.6cm}@{}}
\toprule
\textbf{Pattern} & \textbf{Example} & \textbf{Safety Impact} \\
\midrule
Masked Harassment & \genphrase{Is it acoustic} misclassified & Permits discrimination \\
Context-Dependent & \genphrase{let him cook} misinterpreted & Enables masked bullying \\
Evolution-Based & Delayed \genphrase{skibidi sigma} detection & Allows harmful trends \\
\bottomrule
\end{tabular}
\caption{Critical AI System Failure Patterns}
\label{tab:critical-failures}
\end{table}
\noindent
A further crucial evaluation was about fairness.

Our evaluation revealed significant disparities in AI system interpretation across demographic contexts \cite{gomez_algorithmic_2024}. Variants common among minority youth communities showed 23\% higher false positive rates for policy violations. Context-dependent terms discussing identity and belonging had 31\% higher misclassification rates \cite{dias_oliva_fighting_2021}.
We had the following key findings:
\begin{itemize}
    \item Expression variants common in minority communities: 23\% higher false positives;
    \item Identity and belonging discussions: 31\% higher misclassification;
    \item Harassment detection accuracy: 42\% lower for evolving language targeting marginalized groups.
\end{itemize}
\noindent
These findings align with broader patterns of algorithmic bias in content moderation systems \cite{gallegos_bias_2024} while highlighting specific impacts on Gen Alpha communication practices \cite{dias_oliva_fighting_2021}.

\section{Gen Alpha Vulnerability Landscape}
\label{sec:vulnerability}

Building on the quantitative evaluation results, the next step is the identification of broader patterns of risk and protection failure in Gen Alpha’s digital communication. Section~\ref{sec:vulnerability-assessment} presents a qualitative vulnerability assessment framework that categorizes emerging risks, moderation blind spots, and linguistic ambiguities that complicate safety enforcement. Section~\ref{sec:communication-vulnerability} explores the specific communication dynamics that hinder effective protection, such as platform-specific slang, cross-generational gaps, and help-seeking reluctance.

\subsection{Vulnerability Assessment Framework}
\label{sec:vulnerability-assessment}
Our framework provides a systematic approach to evaluating moderation gaps in Gen Alpha's digital communication, building on established methodologies for analyzing online risks to users \cite{kowalski_electronic_2007}. While traditional frameworks focus on explicit content, our approach specifically addresses the linguistic complexity of Gen Alpha communication \cite{petrova_impact_2021} and the challenges this presents for automated protection systems \cite{hussein_cyberbullying_2023}. The framework is organized along three dimensions, namely, \textit{risk categories} (Section \ref{sec-ra}), \textit{moderation gap metrics } (Section \ref{sec-mgm}) and \textit{help-seeking barriers}, meaning by this barriers preventing the provision of help to Gen Alpha users (Section \ref{sec-hsb}).

\begin{figure}[ht]
    \centering
    \includegraphics[width=0.5\textwidth]{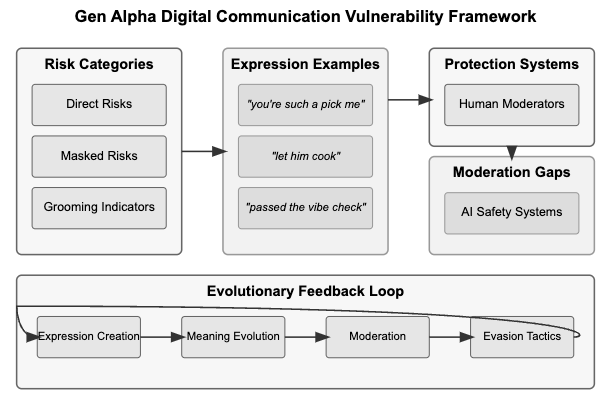}

    \caption{Framework for analyzing Gen Alpha digital communication vulnerabilities. The diagram shows the relationship between different types of risks, protection systems, and resulting moderation gaps.}
    \label{fig:vulnerability-framework}
    \Description{A hierarchical flowchart showing how Gen Alpha digital communication branches into three main risk categories: Direct Risks, Masked Risks, and Grooming Indicators. Direct Risks further divides into Overt Bullying, Direct Exclusion, and Status Attacks. Masked Risks splits into Coded Harassment, Subtle Manipulation, and Social Pressure. Grooming Indicators branches into Trust Building, Isolation Attempts, and Secret-Keeping. All these elements feed into Protection Systems, which splits into two parallel paths: Human Moderators and AI Safety Systems. Both protection system types lead to Moderation Gaps, which ultimately result in three key issues: Communication Barriers, Trust Issues, and System Limitations. Arrows show the flow of influence from each component to the next, illustrating how risks interact with protection mechanisms and where vulnerabilities emerge.}
\end{figure}

\begin{table}[ht]
\begin{tabular}{@{}p{1.2cm}p{1.4cm}p{1.9cm}p{2.6cm}@{}}
\toprule
\textbf{Risk Type} & \textbf{Sub-Category} & \textbf{Example Expression} & \textbf{Content Moderation Implications} \\
\midrule
\multirow{3}{*}{\parbox{1.3cm}{Direct Risks}} 
 & Overt bullying & \genphrase{you're such a pick me} & Explicit social exclusion tactics \\
 & Direct exclusion & \genphrase{NPC behavior} & Dehumanizing language patterns \\
 & Status attacks & \genphrase{beta male behavior} & Hierarchy-based harassment \\
\midrule
\multirow{3}{*}{\parbox{1.3cm}{Masked Risks}} 
 & Coded harassment & \genphrase{is it acoustic} & Discriminatory content masked as inquiry \\
 & Subtle manipulation & \genphrase{let him cook} (mocking) & Context-dependent negativity \\
 & Social pressure & \genphrase{you're not him} & Peer pressure through trending phrases \\
\midrule
\multirow{3}{*}{\parbox{1.3cm}{Grooming Indicators}} 
 & Trust building & \genphrase{you passed the vibe check} & False sense of security creation \\
 & Isolation attempts & \genphrase{they're not giving} & Social separation tactics \\
 & Secret-keeping & \genphrase{keep it lowkey} & Privacy boundary manipulation \\
\bottomrule
\end{tabular}
\caption{Gen Alpha Digital Communication Risk Categories}
\label{tab:risk-categories}
\end{table}

\subsubsection{\textbf{Risk Categories}}
\label{sec-ra}

Figure~\ref{fig:vulnerability-framework} presents an overview of the risk categories and protection systems analyzed, incorporating both established threat models \cite{oksanen_exposure_2014} and emerging challenges specific to Gen Alpha's digital environment \cite{wang_role_2024}. 
Based on the framework in Figure~\ref{fig:vulnerability-framework},
Table \ref{tab:risk-categories} presents the taxonomy of risks identified in Gen Alpha digital communication patterns. This classification has emerged from the analysis of the dataset introduced in Section \ref{sec:dataset}. \textit{Direct risks}, \textit{Masked risks}, and \textit{Grooming risks} organize risks in three categories organized according to how \textit{implicit} the corresponding risks are in a given Gen Alpha expression.
 Each risk type represents distinct challenges for content moderation systems, with examples showing how seemingly innocent phrases can carry significant safety implications in Gen Alpha contexts. 

To operationalize this model in our evaluation and annotation workflow, we applied a structured coding scheme. Table~\ref{tab:risk-coding-framework} presents the taxonomy used by human annotators and expert reviewers during expression labeling. 
Additionally, Table~\ref{tab:evolution-patterns} provides three examples of how common Gen Alpha expressions evolve, most of the time very rapidly, in meaning and intent.
 
\begin{table}[ht]
\centering
\begin{tabular}{|p{2cm}|p{0.7cm}|p{4.4cm}|}
\hline
\textbf{Risk Category} & \textbf{Label Code} & \textbf{Description / Criteria} \\
\hline
Direct Risk & \texttt{R1} & Overtly harmful content such as bullying, threats, or hate speech with explicit negative intent. \\
\hline
Masked Negativity & \texttt{R2} & Harassment or exclusion concealed through slang, memes, or humor; requires contextual inference. \\
\hline
Evolution-Based Risk & \texttt{R3} & Expressions that gain harmful connotation over time or through platform migration. \\
\hline
Low Risk / Neutral & \texttt{R0} & Expressions without significant risk indicators, including benign uses or context-neutral phrases. \\
\hline
Platform-Specific Risk & \texttt{R4} & Terms whose interpretation varies by platform environment (e.g., gaming vs. social media). \\
\hline
Cultural Risk Variant & \texttt{R5} & Risk arising from culture-specific meanings or regional translation effects. \\
\hline
\end{tabular}
\caption{Risk Coding Framework to Evaluate Gen Alpha Expression}
\label{tab:risk-coding-framework}
\end{table}

\begin{table}[ht]
\centering
\begin{tabular}{@{}lll@{}}
\toprule
\textbf{Expression} & \textbf{Original Usage} & \textbf{Evolved Usage} \\
\midrule
\genphrase{Is it acoustic} & Medical inquiry & Discriminatory term \\
\genphrase{Skibidi} & Dance reference & Status marker \\
\genphrase{Rizz} & Charisma & Manipulation marker \\
\bottomrule
\end{tabular}
\caption{Expression Evolution Patterns}
\label{tab:evolution-patterns}
\end{table}

\subsubsection{\textbf{Moderation Gap Metrics}}
\label{sec-mgm}

Moderation effectiveness was measured through several complementary metrics. The pattern detection rate tracked the percentage of concerning expressions successfully identified by both human moderators and AI systems across platforms. Response latency measured the critical time gap between new expression emergence and protection system recognition, revealing significant delays in adapting to evolving language. Context understanding assessed accuracy in interpreting platform-specific meaning variations, particularly in gaming environments where expressions often carried multiple contextual implications. The risk pattern recognition metric evaluated success rates in identifying potentially harmful usage patterns as expressions evolved across different communities and platforms.

\subsubsection{\textbf{Help-Seeking Barriers}}
\label{sec-hsb}

The analysis revealed several interconnected factors that prevented effective protection of Gen Alpha users. Most critically, protection systems consistently lagged behind the rapid evolution of expressions, creating windows of vulnerability where concerning interactions went undetected. This technical gap was compounded by broader context comprehension challenges, where both human moderators and AI systems struggled to understand platform-specific meanings that were intuitively clear to Gen Alpha users. The resulting trust gap led many Gen Alpha users to avoid reporting concerning interactions, believing adults would misunderstand or minimize their experiences. These issues were further exacerbated by detection inconsistency across platforms, where moderation effectiveness varied significantly between different digital spaces that the users regularly navigated.

\subsection{Vulnerability Assessment evaluation}
\label{sec:communication-vulnerability}

Applying the framework in Section \ref{sec:vulnerability-assessment}, we have identified three interconnected challenges compromising the existing protection mechanisms. A first critical vulnerability stems from communication isolation, with 73\% of Gen Alpha users regularly using terms their adult protectors don't understand. This linguistic gap enables concerning patterns across our risk categories:

\begin{itemize}
    \item Platform-specific language masks harmful content (e.g., \genphrase{let him cook} used mockingly);
    \item Gaming terminology obscures subtle harassment (e.g., \genphrase{skill issue});
    \item Trendy expressions facilitate manipulation (e.g., \genphrase{you passed the vibe check}).
\end{itemize}

\noindent
The protection system effectiveness shows concerning limitations across both human and automated monitoring. Humans recognize only 45\% of potentially harmful uses of new expressions; AI systems are even lower performance at 32-42\% accuracy in detecting concerning patterns. This limitation is exacerbated by cross-platform variations that create additional monitoring blind spots. 

\begin{table}[h]
\centering
\begin{tabular}{@{}lll@{}}
\toprule
\textbf{Expression} & \textbf{Positive Usage} & \textbf{Risk Pattern} \\
\midrule
\genphrase{Let him cook} & Performance praise & Sarcastic mockery \\
\genphrase{Skill issue} & Learning opportunity & Harassment marker \\
\genphrase{You're mogging} & Achievement recognition & Status-based exclusion \\
\bottomrule
\end{tabular}
\caption{Gaming Context Expression Analysis}
\label{tab:game-context}
\end{table}

\noindent
To illustrate how identical phrases may function differently across digital environments, Table~\ref{tab:game-context} and Table~\ref{tab:social-context} compare usage and risks in gaming and social settings. 
\begin{table}[h]
\centering
\begin{tabular}{@{}lll@{}}
\toprule
\textbf{Expression} & \textbf{Positive Usage} & \textbf{Risk Pattern} \\
\midrule
\genphrase{Flop era} & Self-deprecation & Group exclusion \\
\genphrase{Main character} & Self-confidence & Mockery target \\
\genphrase{Pick me} & – & Social isolation \\
\bottomrule
\end{tabular}
\caption{Social Media Context Expression Analysis}
\label{tab:social-context}
\end{table}

Perhaps most concerning are the barriers to help-seeking behavior among Gen Alpha users. Our data reveals that the rapid evolution of expressions (averaging 2-3 new meanings per month) consistently outpaces protection systems' ability to adapt. We identified critical protection gaps: 76\% of users showed help-seeking reluctance, 82.5\% of expression meanings evolved faster than moderation updates, and cross-platform monitoring varied by 45\%. This creates a dynamic where harmful interactions can flourish in the gaps between recognized and emerging language patterns. Platform-specific variations further complicate consistent safety monitoring, as expressions may carry different implications across different digital spaces.

When asked to assess vulnerability,  all four LLMs demonstrated relatively high accuracy in identifying direct risks but performed significantly worse on masked risks and evolution-based concerns, see Table \ref{tab:safety-detection}. 
These overall low performance patterns suggest that current AI content moderation systems may offer complementary capabilities to human moderators but cannot fully replace human judgment, particularly for safety-critical decisions involving contextual understanding. 

\begin{table}[ht]
\begin{tabular}{lccc}
\toprule
\textbf{Model} & \textbf{Direct Risks} & \textbf{Masked Risks} & \textbf{Evolution-Based} \\
\midrule
GPT-4 & 72.4 & 45.6 & 38.2 \\
Claude & 75.8 & 48.9 & 41.5 \\
Gemini & 69.3 & 42.8 & 35.7 \\
Llama 3 & 65.7 & 38.4 & 32.3 \\
\bottomrule
\end{tabular}
\caption{Safety Detection Performance by Risk Type (\%)}
\label{tab:safety-detection}
\end{table}

\section{Ethical Considerations}
\label{sec:ethics}

Our analysis revealed several critical insights with significant implications for both research and ethical practice in youth online safety. We found that over two-thirds of concerning interactions used terminology unfamiliar to human moderators, while AI systems failed to detect nearly half of masked harassment attempts. These findings underscore a critical gap that leaves Gen Alpha users particularly vulnerable to sophisticated forms of online manipulation.
The evaluation framework(s) we have proposed demonstrate(s) the importance and feasibility of a youth-centered design if safety systems. By establishing suitable methodologies, this work provides a foundation for future studies while highlighting the unique challenges of conducting research with Gen Alpha participants. These results have 
ethical implications extend in two key directions. 

First, the research process itself requires careful attention to ethical considerations. Thus, for instance, to name the procedures that we enforced in this work:

\begin{itemize}
    \item All participation followed strict consent protocols with both youth and guardian approval;
    \item Data collection focused exclusively on public communications;
    \item Youth participation guidelines were developed and strictly enforced;
    \item Privacy protections were implemented at all research stages.
\end{itemize}
In turn, our findings raise important questions about broader protection concerns in digital spaces, for instance:

\begin{itemize}
    \item How to balance necessary protection with youth autonomy;
    \item Appropriate boundaries for communications monitoring;
    \item Ensuring fairness in automated moderation systems;
    \item Protecting privacy while maintaining safety.
\end{itemize}
These ethical considerations shape not only how such research should be conducted but also how platforms should approach the challenge of protecting Gen Alpha.

As a second key point, this research highlights urgent needs for greater accountability in AI safety systems protecting Gen Alpha users. Current gaps in understanding between AI systems and youth communication demand regular, systematic bias audits that examine system performance across diverse communities and contexts. These audits must go beyond standard metrics to assess how automated moderation affects different segments of the Gen Alpha population.

Meaningful youth consultation represents another, third, critical accountability requirement. Gen Alpha users, particularly those from marginalized communities, must have formal channels to inform both the design and evaluation of safety systems that affect their digital experiences. This consultation should extend beyond superficial feedback to include structured input on how safety systems interpret and moderate their evolving communication patterns.
The appeals process for automated moderation decisions requires particular attention to the Gen Alpha context. Clear, age-appropriate mechanisms must exist for contesting AI decisions, especially given rapid evolution of youth communication patterns. These mechanisms should acknowledge that traditional human moderators may not fully understand the contested communications, necessitating new approaches to appeals review.

Finally, to ensure accountability, platforms must commit to a full documentation of how their automated moderation systems affect different youth communities. This documentation should include:

\begin{itemize}
    \item Regular public reporting on moderation impacts across different demographic groups
    \item Clear disclosure of moderation system limitations regarding youth communication
    \item Transparent metrics on appeals outcomes and response times
    \item Regular updates on efforts to improve system understanding of evolving language patterns
\end{itemize}
These accountability measures aim to ensure that AI safety systems serve their protective function without unduly restricting Gen Alpha's authentic digital expression. The goal is to create a framework where protection and transparency work in concert, rather than opposition.

\section{Conclusion}
\label{sec:conclusion}

This research provides the first systematic evaluation of how AI safety systems interpret Gen Alpha's unique digital communication patterns. By incorporating Gen Alpha users directly in the research process, we've quantified critical comprehension gaps between these young users and their protectors—both human and AI. 

The identification of three critical failure patterns — context-dependent interpretation, masked harassment detection, and evolution-based meaning shifts — highlights specific vulnerabilities in current content moderation approaches. These gaps create dangerous blind spots where concerning interactions may go undetected, particularly given Gen Alpha's documented reluctance to seek adult help when encountering problematic online behavior.

While AI tools can help identify concerning patterns, our findings support Gillespie's argument - full automation of moderation may be neither feasible nor desirable for complex social judgment. Instead, AI systems should enhance human moderation, particularly for Gen Alpha content where context and rapidly evolving meanings are critical. This is the direction of our future work.

\begin{acks}

This research was partially supported by the European Union’s Horizon 2020 FET Proactive project “WeNet — The Internet of Us” (grant agreement no. 823783). The first author conducted this work as a research intern at System Two AI LLC, where she led the data collection, evaluation design, and core authorship of the study. The second author provided mentorship throughout and played a key role in refining the structure, methodology, and academic positioning of the work. The authors also thank Virendra Mehta for coordinating experimental infrastructure and supporting the research process.

\end{acks}

\bibliographystyle{ACM-Reference-Format}
\bibliography{GenAlpha}

\end{document}